\documentclass[showpacs,amsmath,amssymb,twocolumn,floatfix,prl]{revtex4}
\usepackage[dvips]{graphicx}
\input{epsf}

\begin{document}

\title{Directing transport by polarized radiation
in presence of chaos and dissipation}


\author{A.D.~Chepelianskii$^{(a)}$ and 
D.L.~Shepelyansky$^{(b)}$}
\homepage[]{http://w3-phystheo.ups-tlse.fr/~dima}
\affiliation{$^{(a)}$Ecole Normale Sup\'erieure, 45, rue d'Ulm, 
75231 Paris Cedex 05, France\\
$^{(b)}$Laboratoire de Physique Th\'eorique, 
UMR 5152 du CNRS, Universit\'e P. Sabatier, 31062 Toulouse Cedex 4, France
}

\date{February 27, 2004}

\begin{abstract}
We study numerically the dynamics of particles on the Galton board of semi-disk
scatters in presence of monochromatic radiation and dissipation. It is shown
that under certain conditions the radiation leads to appearance
of directed transport linked to an underlining strange attractor.
The direction of transport can be efficiently changed by
radiation polarization. The experimental realization
of this effect in asymmetric antidot superlattices  is discussed.

\end{abstract}


\maketitle
A great challenge for the future technology on microscopic scale
is the ability to control transport 
created by external energy sources in presence of dissipation
and noise. These sources may drive a system out of equilibrium
and generate a transport which direction is related
to the system configuration in a rather nontrivial way. 
Such a directed transport appears in  Brownian motors, 
or ratchets, which now attract a great interest 
of the community \cite{hanggi,reimann,prost}.
At present the technological progress allowed to 
observe ratchets experimentally in a variety of
systems including semiconductor heterostructures
\cite{linke}, vortices in Josephson junction arrays 
\cite{mooij,nori}, cold atoms \cite{grynberg},
macroporous silicon membranes \cite{muller} and other systems.
Biological applications of Brownian motors are also of
primary importance \cite{hanggi,reimann,prost}.

The theoretical studies of ratchets have led to 
a number of interesting results (see \cite{reimann}
and Refs. therein) including such unexpected phenomena
as absolute negative mobility when the transport
is directed against an applied force \cite{hanggi1}.
However, up to now the studies of ratchets
have been done mainly \cite{note} in the limit of strong
dissipation neglecting the effects of second time derivative
\cite{hanggi,reimann} or, in opposite, only the Hamiltonian
dynamics has be analyzed \cite{flach,roland}. 
In this Letter we consider an intermediate case with a moderate
or weak dissipation in a two-dimensional dynamical system
driven by monochromatic polarized radiation. 
Without dissipation the dynamics
is completely chaotic and is characterized by a
homogeneous diffusion. 
Surprisingly, the dissipation related to friction
leads to appearance of a directed transport 
which direction can be changed efficiently by
polarization of radiation. 

Our dynamical model describes the motion of noninteracting particles
on the modified Galton board in presence of friction and  polarized
monochromatic radiation. The board of
rigid disks on a triangular lattice had been introduced by 
Galton in 1889 to analyze the dynamics of 
particles elastically colliding with disks \cite{galton}.
In absence of friction and radiation the dynamics is completely
chaotic as it had been proven by Sinai (see {\it e.g.} \cite{sinai}).
To break the symmetry of the Galton board we replace disks by semi-disks
oriented in the $x$ direction as shown in Fig.~1. 
Such a system has only translation and reflection $y \rightarrow -y$ 
symmetries. 

\begin{figure}[t!]  
\centerline{\epsfxsize=8.5cm\epsffile{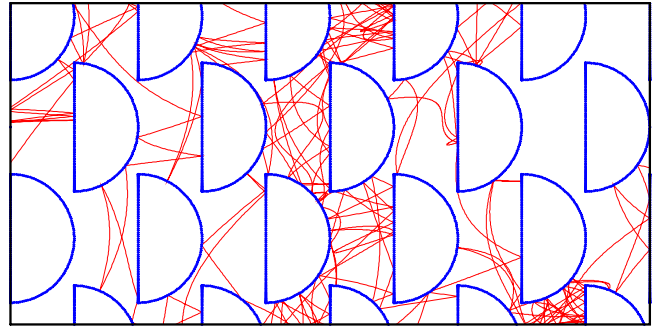}}
\vglue -0.0cm 
\caption{(color online) A chaotic trajectory 
(red/grey curve)
on the Galton board of semi-disks (blue/black) at radiation with
$f=4$, $\theta=\pi/8$, $\omega=1.5$ and dissipation $\gamma=0.1$.
The trajectory starts near $x=y=0$,
the region shown corresponds to $-27<x<-17$, $-10<y<-5$.
Same trajectory on a large scale is shown in Fig.3.
}
\label{fig1}       
\end{figure}

We analyze the dynamics of particles on this modified Galton board
under the influence of a linearly polarized radiation which gives a
force directed by angle $\theta$ to  $x$-axis: 
$\mathbf{f} = f  (\cos \theta, \sin \theta) \cos{\omega t}$. 
In addition a particle experiences a friction force 
$\mathbf{F}_f = - \gamma \mathbf{v}$ directed against its velocity,
and elastic collisions with semi-disks. 
A typical example of a chaotic trajectory 
in this regime is shown in Fig.~1. 
Without friction the dynamics 
is chaotic and a diffusive spreading in $x$ and $y$ directions
takes place as well as a diffusive energy growth induced by {\it ac}-driving. 
However, due to the time reversibility, a  directed transport is not possible 
in absence of friction. The introduction of friction breaks 
the time inversion symmetry and creates a directed transport. 

To study numerically the dynamics of this model, we fix the disk radius
$r_d=1$ and the particle mass $m=1$ so that the system is characterized 
only by the distance $R$ between disk centers. In the following
we consider the case of compact semi-disk board with $R=2$
(a moderate variation of $R$ does not change qualitatively the 
dynamics properties). Following the approach of \cite{first},
the particle dynamics is simulated numerically by using the exact 
solution of Newton equations between collisions, and by determining the
collision points with the rapidly converging Newton algorithm.
This way at $f=0, \gamma=0$ the total energy is conserved with 
a relative precision of $10^{-14}$. A typical example of an average
density distribution obtained from a simulation of many trajectories
is shown in Fig.~2. In spite of chaos the dissipation leads to an
average directed transport of particles moving under angle $\psi \approx 2.3$
to $x$-axis.

\begin{figure}[t!]  
\centerline{\epsfxsize=8.5cm\epsffile{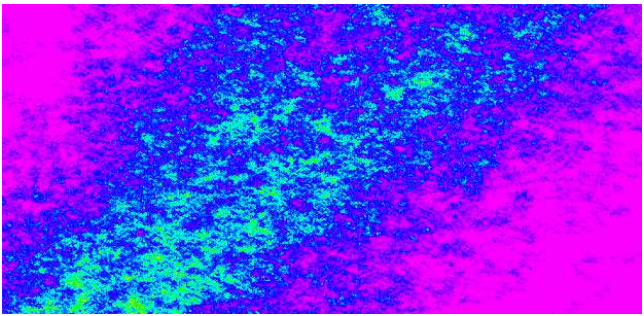}}
\vglue -0.0cm 
\caption{(color online) Density distribution in the region
$-400 \leq x \leq -100$, $0 \le y \leq 600$ averaged over
200 trajectories in a time interval $0 \leq t \leq 2 \cdot 10^4$
for parameters of Fig.~1. Initially all trajectories are 
distributed near $x=y=0$ with random velocity directions
and $v^2 = 1$. For convenience the figure is clockwise
rotated on $\pi/2$, density is proportional to color
changing from zero (rose/black) to maximum (green/white).
}
\label{fig2}       
\end{figure}

The direction of transport depends on the polarization of radiation
as it is shown in Fig.~3 for typical trajectories computed 
on large time and space scales. 
The  average velocity of directed flow can be written as
${\mathbf v_f} = v_f (\cos \psi, \sin \psi)$.
For radiation polarized
along $x$ ($\theta=0$)
the average transport goes along $-x$ direction ($\psi=\pi$)
while for  radiation polarized along $y$ ($\theta=\pi/2$)
the direction of transport is inversed ($\psi=0$).
The fact that the particles move along $x$-axis for these $\theta$
values is in agreement with the symmetry properties of the 
semi-disks board. However, in general the flow direction depends
in a nontrivial way on the chaotic dynamics 
in presence of radiation and dissipation \cite{note1}.
By changing the polarization angle $\theta$ between 
$0$ and $\pi/2$ it is possible to
change smoothly the angle of directed flow $\psi$  from $\pi$ to $0$
(see Figs.~3,4). The values of
$v_f, \psi$ and average velocity  square $\; v^2 = \; <v_x^2+v_y^2>$
are determined from  one long
trajectory (up to times with $\omega t \sim 10^7$) or from 
averaging over 10 shorter orbits.
The functional dependence $\psi(\theta)$ found 
in this way is shown in Fig.~4.
It varies moderately with system parameters 
but,  in average,  can be approximated by a linear relation
$\psi=\pi - 2 \theta$. We note that in presence of radiation
the directionality properties of transport are much more
rich compared to the case of static applied force where for disks
the chaotic transport simply follows the force direction 
\cite{first,hoover}.

\begin{figure}[t!]
\centerline{\epsfxsize=8.5cm\epsffile{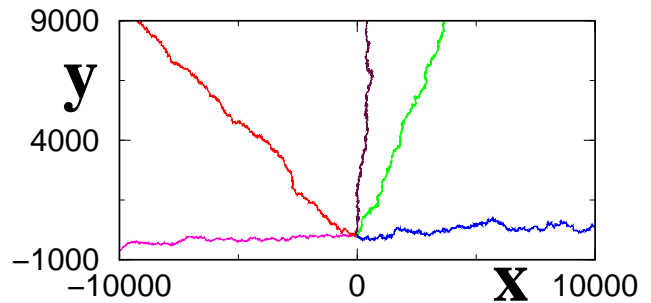}}
\vglue -0.0cm
\caption{(color online) Directed transport 
for one trajectory at various polarizations of radiation 
$\theta = 0$, $\pi/8$ (same trajectory as in Fig.~1),
$0.21 \pi$, $\pi/4$, $\pi/2$ (from left to right clockwise)
at $f=4$, $\omega=1.5$, $\gamma=0.1$.
}
\label{fig3}
\end{figure}

To understand the properties of the directed transport
let us start from the analysis of its averaged characteristics.
It is clear that in presence of chaos 
{\it ac}-driving leads to an effective heating increasing 
the particle kinetic energy $E = v^2/2$. This growth is stopped
by energy dissipation induced by friction. The balance of these two processes
gives an average energy $E$ of motion which due to chaos and dissipation
takes place on a strange attractor \cite{ott}. This energy can be find on 
the basis of following estimates. Indeed, from the Newton equations of
motion: $\dot{E} \sim v \dot{v} \sim f v \cos \omega t$.
Hence, the diffusion rate in energy $D_E = (\Delta E)^2/t \sim
{\dot{E}}^2 \tau \sim f^2 v l$ where the mean-free path $l \sim R \sim 1$
determines the collision time $\tau = l/v \sim R/E^{1/2}$
and the space diffusion rate $D \sim v l$.
Let us assume that the friction is weak and many collisions take
place during the  dissipative time scale $1/\gamma$. Then,
at the equilibrium the diffusion in energy is compensated by dissipation
so that the average energy is determined in a self-consistent
way as 
\begin{equation}
\label{velav}
E=v^2/2 \sim (D_E/\gamma)^{1/2} 
\sim  ( l f^2/ \gamma )^{2/3} \sim (f^2/\gamma)^{2/3}\;\; 
\end{equation}
where in the last relation we used that in our case $l \sim R \sim 1$.
The same expression can be also obtained in a more formal way by
writing the relation for velocities $v_n, v_{n+1}$ between two consecutive
collisions $n$ and $n+1$  from exact solution of the Newton equations.
Then, assuming that the dynamics is ergodic and chaotic
on the energy surface, the squares of these velocities may 
be averaged over all directions and times between collisions.
In the equilibrium both average $v_n^2$ and $v_{n+1}^2$ are equal
that gives the relation (\ref{velav}). The derivation of this expression
assumes that many collisions take place. This requires that the amplitude
of oscillations induced by radiation $a$ is larger then the distance
between semi-disks: $a=f/(m \omega^2) \gg R$. 
Otherwise a trajectory drops on a simple attractor without
scattering on semi-disks.

The numerical data are presented in Fig.~5 and confirm 
the dependence (\ref{velav})
in almost 4 orders of magnitude range. The deviations appear only at 
large values of $\gamma$ when the dissipation rate becomes
comparable with the frequency of collisions.
It is also interesting to note that in agreement with
(\ref{velav}) the dependence on radiation frequency $\omega$ is rather weak.
Indeed, a change of $\omega$ by an order of magnitude 
gives only a factor of 2 change in $v^2$ and 
in the flow velocity $v_f$ (see Figs.~5,6). 
Even a larger change  from $\omega=2.5$ to $0.025$ at $f=4$, $\gamma=0.1$
gives only $30\%$ increase of $v_f$ and $50\%$ drop of $v^2$ (data not shown).
Such a weak dependence on $\omega$ can be attributed to 
finite time correlations, existing in the chaotic dynamics,
which have been neglected in the derivation of (\ref{velav}).
The directionality of the flow is globally robust in respect to  
parameter variation (see Fig.~4). However,  the 
dependence $\psi(\theta)$ has certain local changes
for small $\theta$ when $\omega$ becomes smaller or comparable
with $\gamma$. 

\begin{figure}[t!]
\epsfxsize=3.2in
\epsffile{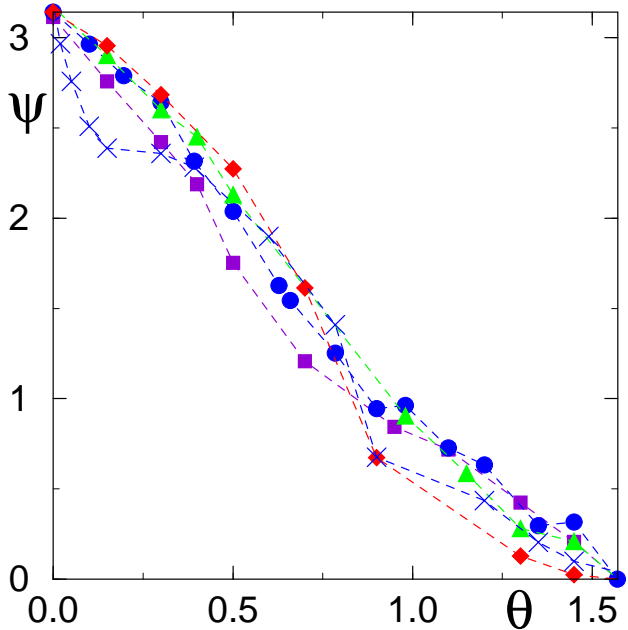}
\vglue -0.0cm
\caption{(color online) Dependence of flow angle $\psi$ on 
polarization angle $\theta$, measured in radians, for $f=4$ and $\omega=1.5$
at $\gamma =$ 0.2 (squares), 0.1 (circles), 0.05 (triangles)
and 0.025 (diamonds) and $\omega=0.1$
at $\gamma=0.1$ ($\times$). Dashed curves are drown to adapt an eye.
}
\label{fig4}       
\end{figure}

\vglue 0.0cm
\begin{figure}
\epsfxsize=3.2in
\epsfysize=3.2in
\epsffile{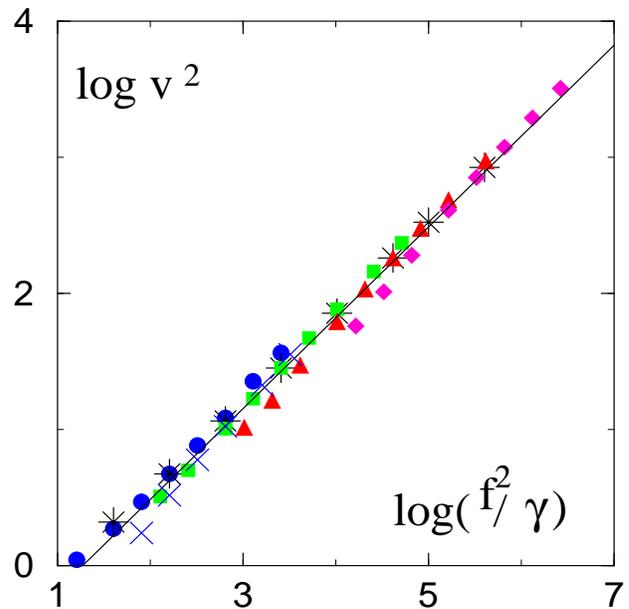}
\vglue -0.0cm
\caption{(color online) Dependence of average $v^2$ on $f$
and $\gamma$ for  $\theta=0$ at $\omega=1.5$
and $f = $4(circles), 16 (squares), 64 (triangles), 
256 (diamonds)  and 
$\gamma = 0.1$ ($*$); the case at $\omega=0.1$ and
$f=4$ is shown by ($\times$);
total interval of parameters variation is
$5 \cdot 10^{-3} \leq \gamma \leq 4$,
$2 \leq f \leq 256$. The full line gives the dependence
$v^2=(f^2/\gamma)^{2/3}/7$. Logarithms are decimal here and in Fig.~6.
}
\label{fig5}       
\end{figure}

The dependence of the velocity $v_f$ of the directed flow
${\mathbf v_f}$ is shown in Fig.~6 for $\theta=0$.
The numerical data are well fitted by the relation:
\begin{equation}
\label{veflow}
v_f \approx l^{2/3} (f \gamma/m^2)^{1/3}/12 \; (\gamma > \gamma_c)\; ;  \;
v_f \approx l \gamma/m \; (\gamma < \gamma_c) \; 
\end{equation}
where $\gamma_c \approx (m f/l)^{1/2}/40$ and we assumed $m=l=r_d=1$.
The dependence of $v_f$ on parameters for $ \gamma > \gamma_c$
reminds the situation for a case of the Galton board with disks
in presence of static force and friction (see \cite{first,hoover}).
Indeed, assuming that the flow moves in a given direction
its velocity can be estimated as $v_f \sim f \tau_c/m$
where $\tau_c = l/v$. According to (\ref{velav}) this leads
to (\ref{veflow}) for $ \gamma > \gamma_c$.
However, in a difference from \cite{first,hoover} 
the dependence is changed in the limit of small $\gamma \ll \gamma_c$.
Qualitatively, one may say that in this limit the particle velocity
becomes very large, the chaotic collisions become very often
and this leads to a significant averaging and decrease of $v_f$,
which becomes proportional to the dissipation $\gamma$ that is responsible
for appearance of directed flow. With the variation of $\theta$
from $0$ to $\pi/2$ the value of $v_f$ drops approximately by 
a factor 2 ({\it e.g.} for data in Fig.~4).

The results presented above show that the appearance of directed
transport induced by polarized radiation is a robust phenomenon
which is not very sensitive to the parameters of the model.
Unfortunately we are not able to derive analytically the 
directionality dependence $\psi(\theta)$. Qualitatively, we may argue that
for $\theta=0$ the dissipation leads to a trapping of particles
in a funnel formed by semi-disks and the flow goes to the left.
On the contrary, for $\theta=\pi/2$ the vertical oscillations
induced by radiation push particles out of a funnel
and in presence of dissipation the flow goes to the right.
This indicates that approximately $\psi = \pi - 2\theta$
in agreement with the numerical data. However, it is desirable
to  have a more rigorous derivation in addition to 
these handwaving arguments. We also note that in the case
of charged particles an introduction of a magnetic field
leads to a reorientation of the directed flow following approximately
the direction given by the average Lorentz force
(we leave a discussion of this effect for 
further studies).

\vglue 0.0cm
\begin{figure}
\epsfxsize=3.2in
\epsfysize=3.2in
\epsffile{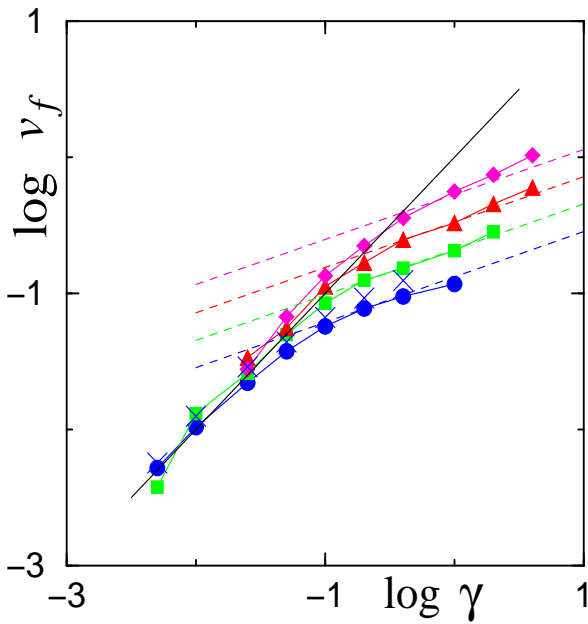}
\vglue -0.0cm
\caption{(color online) Dependence of the flow velocity
$v_f$ on friction $\gamma$ for  $\theta=0$ and
$\omega=1.5$ at $f = $ 4,
16, 64 and 256 
and $\omega=0.1$ at $f = $ 4 (same symbols as in Fig.~5).
The straight lines show the asymptotic behavior (\ref{veflow})
at $\gamma > \gamma_c$ (dashed) and $\gamma < \gamma_c$ (full).
}
\label{fig6}       
\end{figure}

The directed transport induced by polarized radiation
demonstrates rather interesting and unusual properties
and it would be interesting to study it in experiments
with semiconductor heterostructures. The nowadays technology allows
to realize experimentally antidot superlattices
(see {\it e.g.} \cite{weiss}) and it is known that
classical chaotic dynamics plays an important role
in transport properties of such systems (see {\it e.g.} \cite{geisel}).
However, up to now only circular antidots have been studied. 
In fact technologically it is rather simple
to replace disks by semi-disks and to place the whole board
in a microwave field. Moreover, the effects of microwave
radiation on transport in disordered mesoscopic systems
have been studied experimentally \cite{kvon} and it was shown
that radiation may generate a directed transport 
in a case of broken space symmetry. Since disorder is
symmetric in average it is clear that the ratchet effect 
should be significantly larger in artificially prepared asymmetric
structures as the Galton board of semi-disks  discussed here.
A typical set of experimental parameters can be 
$r_d \sim R \sim l \sim 1 \mu$m, $\omega/2\pi = 1$GHz, $m=0.067m_e$.
For a microwave field of strength $\epsilon = 10$ V/cm
the amplitude of particle oscillations is
$a = e \epsilon /m \omega^2  \approx 60 \mu$m ($a \gg R$), so that
the appearance of a simple attractor is excluded
and a directed transport should appear due to presence of
dissipative processes. A possibility to study transport in 
antidot lattices in presence of microwave radiation
has been experimentally demonstrated in \cite{kvon1}
and therefor the experimental realization of our model
should be possible. Such experiments 
should allow to direct transport by radiation on mesoscopic scale
and to study the dissipative processes in such systems 
from a new view point. The theoretical model discussed here should
be also further developed to be more adapted for such experiments
({\it e.g.} Fermi surface effects and more realistic dissipation
should be taken into account).

We thank Kvon Ze Don for useful discussions.


\begin{thebibliography}{99}
\bibitem{note1} In shops one can find a children toy board,
     placed between two plastic glasses,
     with disks replaced by same way oriented triangles
     and metallic balls moving on a board; shaking the board
     in $x$-axis moves all balls from right side of the board
     to the left. This corresponds to the case $\theta=0$ in Fig.~3.
\end{thebibliography}
\end{document}